\documentclass[11pt]{article}

\usepackage{epsfig}
\usepackage{latexsym}
\usepackage{graphicx}
\usepackage{color}
\usepackage{amsmath,amssymb}
\usepackage{cite}
\usepackage{cancel}
 \usepackage{pstricks}
\usepackage{afterpage}
\usepackage{framed}
\usepackage{multirow}
\usepackage{seqsplit}
\usepackage{enumitem}

\usepackage{amsmath, amsthm, amsfonts, amssymb, latexsym}
\usepackage{upgreek}
\usepackage[caption=false]{subfig}

\usepackage{url}
\usepackage{ifpdf}

\def\be{\begin{equation}}
\def\ee{\end{equation}}
\def\ba{\begin{eqnarray}}
\def\ea{\end{eqnarray}}
\def\ra{\rightarrow}

\def\ltap{\;\centeron{\raise.35ex\hbox{$<$}}{\lower.65ex\hbox{$\sim$}}\;}
\def\gtap{\;\centeron{\raise.35ex\hbox{$>$}}{\lower.65ex\hbox{$\sim$}}\;}

\newcommand{\bea}{\begin{eqnarray}}
\newcommand{\eea}{\end{eqnarray}}

\makeatletter
\def\section{\@startsection {section}{1}{\z@}{-3.5ex plus -1ex minus -.2ex}{2.3ex plus .2ex}{\large\bf}}
\def\subsection{\@startsection{subsection}{2}{\z@}{-3.25ex plus -1ex
minus -.2ex}{1.5ex plus .2ex}{\normalsize\bf}}
\makeatother
\newcommand{\captionfonts}{\small}
\makeatletter  % Allow the use of @ in command names
\long\def\@makecaption#1#2{%
  \vskip\abovecaptionskip
  \sbox\@tempboxa{{\captionfonts #1: #2}}%
  \ifdim \wd\@tempboxa >\hsize
    {\captionfonts #1: #2\par}
  \else
    \hbox to\hsize{\hfil\box\@tempboxa\hfil}%
  \fi
  \vskip\belowcaptionskip}
\makeatother   % Cancel the effect of \makeatletter
\catcode`@=11
\def\marginnote#1{}
\newcount\hour
\newcount\minute
\newtoks\amorpm
\hour=\time\divide\hour
by60
\minute=\time{\multiply\hour by60 \global\advance\minute
by-\hour}
\edef\standardtime{{\ifnum\hour<12 \global\amorpm={am}
\else\global\amorpm={pm}\advance\hour by-12 \fi
 \ifnum\hour=0
\hour=12 \fi
 \number\hour:\ifnum\minute<10
0\fi\number\minute\the\amorpm}}
\edef\militarytime{\number\hour:\ifnum\minute<10
0\fi\number\minute}
\def\draftlabel#1{{\@bsphack\if@filesw
{\let\thepage\relax
 \xdef\@gtempa{\write\@auxout{\string
\newlabel{#1}{{\@currentlabel}{\thepage}}}}}\@gtempa
 \if@nobreak
\ifvmode\nobreak\fi\fi\fi\@esphack}
\gdef\@eqnlabel{#1}}
\def\@eqnlabel{}
\def\@vacuum{}
\def\draftmarginnote#1{\marginpar{\raggedright\scriptsize\tt#1}}
\def\draft{\oddsidemargin
0.0truein
 \def\@oddfoot{\sl preliminary draft \hfil
\rm\thepage\hfil\sl\today\quad\militarytime}
 \let\@evenfoot\@oddfoot
\overfullrule 3pt
 \let\label=\draftlabel
\let\marginnote=\draftmarginnote
\def\@eqnnum{(\theequation)\rlap{\kern\marginparsep\tt\@eqnlabel}
\global\let\@eqnlabel\@vacuum}
}
\catcode`@=12
\newcommand{\beq}{\begin{eqnarray}}
\newcommand{\eeq}{\end{eqnarray}}

\begin{document}

\begin{flushright}
CFTP/15-006
\end{flushright}

%\preprint{CFTP/15-003}
\thispagestyle{empty}

\begin{center}
%\hfill \\

\begin{center}

\vspace{1.7cm}

%{\LARGE\bf Conclusive signals of CP-violation in Higgs decays at the LHC run 2}
%\end{center}

{\LARGE\bf Undoubtable signs of CP-violation in Higgs decays at the LHC run 2}
\end{center}

\vspace{1.4cm}

\renewcommand{\thefootnote}{\fnsymbol{footnote}}
{\bf Duarte Fontes$^{\,1\,}$}\footnote{E-mail: \texttt{duartefontes@tecnico.ulisboa.pt}},
{\bf Jorge C. Rom\~ao$^{\,1}\,$}\footnote{E-mail: \texttt{jorge.romao@tecnico.ulisboa.pt}},
{\bf Rui Santos$^{\,2,\,3\,}$}\footnote{E-mail: \texttt{rasantos@fc.ul.pt}} and
{\bf Jo\~{a}o P.~Silva$^{\,1\,}$}\footnote{E-mail: \texttt{jpsilva@cftp.ist.utl.pt}}
\\

\vspace{1.cm}

${}^1\!\!$
{\em Departamento de F\'{\i}sica and CFTP, Instituto Superior T\'{e}cnico, Universidade de Lisboa,} \\
{\em Avenida Rovisco Pais 1, 1049-001 Lisboa, Portugal}
\\
${}^2\!\!$
{\em Centro de F\'{\i}sica Te\'{o}rica e Computacional,
    Faculdade de Ci\^{e}ncias,
    Universidade de Lisboa,} \\
{\em Campo Grande, Edif\'{\i}cio C8 1749-016 Lisboa, Portugal}
\\
${}^3\!\!$
{\em {ISEL -
 Instituto Superior de Engenharia de Lisboa,\\
 Instituto Polit\'ecnico de Lisboa
 1959-007 Lisboa, Portugal}
}\\

\end{center}

\vspace{1.8cm}
\centerline{\bf Abstract}
\vspace{2 mm}
\begin{quote}
\small With the discovery of the Higgs boson at the Large Hadron Collider
the high energy physics community's attention has now turned to understanding the properties
of the Higgs boson, together with the hope of finding more scalars during run 2.
In this work we discuss scenarios where using a combination of three decays,
involving the 125 GeV Higgs boson, the Z boson and at least one more scalar,
an indisputable signal of CP-violation arises.

We use a complex two-Higgs doublet model as a reference
model and present some benchmark points that have passed all
current experimental and theoretical
constraints, and that have cross sections large enough to be probed during run 2.
\end{quote}

\newpage
\setcounter{page}{1}
\setcounter{footnote}{0}

%%%%%%%%%%%%%%%%%%%%%%%%%%%%%%%%%%%%%%%%%%%%%%%%%%%%%%%
\section{\label{sec:intro} Introduction}

The discovery of the Higgs boson by the ATLAS~\cite{ATLASHiggs} and CMS~\cite{CMSHiggs} collaborations
at the Large Hadron Collider (LHC) has raised the interest of the
high energy physics community
in multi-scalar models. One of the most attractive features of some of these models is to provide
extra sources of CP-violation which could help to explain the matter anti-matter asymmetry of the Universe.
This was the reason that lead T.D.~Lee to propose the two-Higgs double model (2HDM)~\cite{Lee:1973iz}
as a means to explain this asymmetry.
Reviews of the 2HDM may be found, for example, in \cite{hhg, ourreview}.
One of the CP-violating complex versions of the 2HDM, which we refer to as
C2HDM, has been the subject of many recent studies~\cite{Barroso:2012wz, Inoue:2014nva,Cheung:2014oaa,
Fontes:2014xva, Fontes:2015mea, Chen:2015gaa, Fontes:2015gxa}. The C2HDM was first proposed 
in~\cite{Ginzburg:2002wt} and it is the simplest version of an explicit CP-violating 2HDM
with a clear and easy limit leading to its CP-conserving version.

As proposed in \cite{BLS},
CP-violation in the scalar sector can be found in the interactions with
gauge bosons in a very simple way.
If CP were conserved,
any decay $h_i \rightarrow h_j Z$ would imply opposite
CP parities for $h_i$ and $h_j$.
Moreover,
assuming only lagrangian terms up to dimension four, any scalar $h_i$
decaying into $ZZ$ would be
CP even~\footnote{There are CP conserving terms of
dimension higher than four that can mediate the decay of a
pseudoscalar into two vector bosons. Those could appear at
loop level from a fundamental theory, but would lead to rates far smaller
than the tree level rates considered in this article. A calculation performed 
in the framework of the 2HDM has shown~\cite{Arhrib:2006rx} that the 
loop mediated decays of the type $h_i \to ZZ$ are several orders of magnitude smaller
than the tree-level ones.}.
Thus,
for example, the simultaneous presence of the decays
$h_3 \rightarrow h_2 Z$,
$h_2 \rightarrow h_1 Z$,
and $h_3 \rightarrow h_1 Z$
violates CP.
We say that points in the C2HDM parameter space which lead to this situation
belong to class $C_1$.
Similarly (with the caveat in footnote 1),
the simultaneous presence of the decays
$h_i \rightarrow h_j Z$,
$h_i \rightarrow Z Z$,
and $h_j \rightarrow Z Z$,
also violates CP.
Within the 2HDM,
there are three such possibilities, according
to the $(i,j)$ assignments,
which we name classes $C_2$, $C_3$, and $C_4$.
Notice that classes $C_1$-$C_4$ represent CP-violation,
regardless of the origin of the neutral scalars.
They may come from an $N$ Higgs doublet model,
or indeed from scalar fields in any number and
from any representation of $SU(2)_L$
(singlets, doublets, triplets, combinations thereof, etc\dots)
In Table~\ref{tab:classes},
\begin{table}[!h]
 \small
 \centering
\begin{tabular}{|c|l|l|l|l|l|}\hline
Classes & $C_1$ & $C_2$ & $C_3$ & $C_4$ & $C_5$ \\\hline
        & $h_3\to h_2 Z$ & $h_2\to h_1 Z$ & $h_3\to h_1 Z$ & $h_3\to h_2 Z$
        & $h_3\to Z Z$ \\
Decays  & $h_2\to h_1 Z$ & $h_1\to Z Z$   & $h_1\to Z Z$   & $h_2\to Z Z$
        & $h_2\to Z Z$ \\
        & $h_3\to h_1 Z$ & $h_2\to Z Z$   & $h_3\to Z Z$   & $h_3\to Z Z$
        & $h_1\to Z Z$ \\
 \hline \end{tabular}
 \caption{\label{tab:classes}Classes of combined measurements guaranteed
 to probe CP-violation in 2HDMs.}
\end{table}
we show the decays involved in each class.
Furthermore,
in the specific context of a 2HDM,
the properties of the fields ensure that,
if CP were conserved,
there would be two CP even neutral scalars
and one CP odd neutral scalar,
usually denoted by $H$, $h$, and $A$,
respectively.
Thus,
in the 2HDM,
the simultaneous presence of
$h_i \rightarrow ZZ$ for $i=1,2,3$ signals CP-violation.
We denote that possibility by class $C_5$.
We stress that class $C_5$ does not represent
necessarily CP-violation in models other than the 2HDM.
For example, even with three Higgs doublets one will surely
have three neutral scalars and class $C_5$ would be consistent with
CP-conservation. We will further discuss other classes that probe
CP-violation that involve one scalar to two scalar decays that
usually have the drawback of having smaller cross sections.

It is interesting that there are only three basis-invariant quantities
signalling CP-violation in the scalar sector of the 2HDM.
They were introduced in \cite{LS, BS},
the connection with the observables explained in \cite{BLS},
and revisited in \cite{Grzadkowski:2014ada}.
Measurements of classes $C_1$-$C_5$ are enough to probe
all invariants.
In the particular setting of the C2HDM,
there is only one phase/source of CP-violation,
all invariants are related,
and the CP-violation in all classes 
(which one can take as the product of the three rates in each class)
is proportional to that phase.

One of the most interesting points of our proposal is that although
the above described classes constitute an indisputable sign of CP-violation,
they have all been searched for individually at run 1. In fact, the searches
$h_i \to ZZ$ and $h_i \to h_j Z$ were already performed by both the ATLAS
and CMS collaborations. Therefore, as long as we have enough signal events
in three of the proposed channels for a given set of parameters, there are
good chances of observing direct CP-violation at the next LHC run.  

This paper is organized as follows. In section~\ref{sec:model}, we briefly describe
the complex 2HDM and the theoretical and phenomenological constraints imposed
on the model with special emphasis on the most recent LHC data. 
In section~\ref{sec:bench}, we propose a set of CP-violating benchmarks points
for Type II and for the Flipped model. In the same section we discuss clear signs
of CP-violation that involve the decay of one scalar to two scalars.
Our conclusions are presented in Section~\ref{sec:concl}.
Finally, we present benchmark points for Type I and for the Lepton Specific model
in appendix~\ref{sec:app}.

%%%%%%%%%%%%%%%%%%%%%%%%%%%%%%%%%%%%%%%%%%%%%%%%%%%%%%%
\section{\label{sec:cxSM} The complex two-Higgs doublet model}
\label{sec:model}

We use as a benchmark model an extension of the SM with an extra scalar doublet.
This complex 2HDM has a softly broken $Z_2$ symmetry $\phi_1 \ra \phi_1, \phi_2 \ra -\phi_2$
and the scalar potential is written as~\cite{ourreview}
\ba
V_H
&=&
m_{11}^2 |\phi_1|^2
+ m_{22}^2 |\phi_2|^2
- m_{12}^2\, \phi_1^\dagger \phi_2
- (m_{12}^2)^\ast\, \phi_2^\dagger \phi_1
\nonumber\\*[2mm]
&&
+\, \frac{\lambda_1}{2} |\phi_1|^4
+ \frac{\lambda_2}{2} |\phi_2|^4
+ \lambda_3 |\phi_1|^2 |\phi_2|^2
+ \lambda_4\, (\phi_1^\dagger \phi_2)\, (\phi_2^\dagger \phi_1)
\nonumber\\*[2mm]
&&
+\, \frac{\lambda_5}{2} (\phi_1^\dagger \phi_2)^2
+ \frac{\lambda_5^\ast}{2} (\phi_2^\dagger \phi_1)^2 \, ,
\label{VH}
\ea
and because the potential has to be hermitian, all couplings except  $m_{12}^2$ and $\lambda_5$ are real.
In order to assure that the two phases cannot be removed simultaneously, we impose
 $\textrm{arg}(\lambda_5) \neq 2\, \textrm{arg}(m_{12}^2)$~\cite{Ginzburg:2002wt}.
By taking $m_{12}^2$ and $\lambda_5$ real we recover the corresponding CP-conserving 2HDM.

The model has three neutral particles with no definite CP, $h_1$, $h_2$ and $h_3$,
and two charged scalars $H^\pm$. The mass matrix of the neutral scalar states
is obtained via the rotation matrix~\cite{ElKaffas:2007rq}
\be
R =
\left(
\begin{array}{ccc}
c_1 c_2 & s_1 c_2 & s_2\\
-(c_1 s_2 s_3 + s_1 c_3) & c_1 c_3 - s_1 s_2 s_3  & c_2 s_3\\
- c_1 s_2 c_3 + s_1 s_3 & -(c_1 s_3 + s_1 s_2 c_3) & c_2 c_3
\end{array}
\right)
\label{matrixR}
\ee
with $s_i = \sin{\alpha_i}$ and
$c_i = \cos{\alpha_i}$ ($i = 1, 2, 3$) and
\be
- \pi/2 < \alpha_1 \leq \pi/2,
\hspace{5ex}
- \pi/2 < \alpha_2 \leq \pi/2,
\hspace{5ex}
- \pi/2 \leq \alpha_3 \leq \pi/2.
\label{range_alpha}
\ee

The C2HDM has 9 independent parameters which we choose to be $v$, $\tan \beta$, $m_{H^\pm}$,
$\alpha_1$, $\alpha_2$, $\alpha_3$, $m_1$, $m_2$, and $\textrm{Re}(m_{12}^2)$.
With this choice the mass of heavier neutral scalar is a dependent parameter given by
\be
m_3^2 = \frac{m_1^2\, R_{13} (R_{12} \tan{\beta} - R_{11})
+ m_2^2\ R_{23} (R_{22} \tan{\beta} - R_{21})}{R_{33} (R_{31} - R_{32} \tan{\beta})}.
\label{m3_derived}
\ee
and the parameter space will be restricted to values which obey $m_3 > m_2$.

We will analyse the usual four Yukawa versions of the C2HDM, in which the $Z_2$ symmetry is extended
to the Yukawa Lagrangian~\cite{GWP} in order to avoid flavour changing neutral currents (FCNC).
In all models the up-type quarks couple to $\phi_2$ and
the so-called Type I (Type II) is obtained by coupling down-type quarks and charged leptons
to $\phi_2$ ($\phi_1$), while by coupling
the down-type quarks to $\phi_1$ and the charged leptons
to $\phi_2$ we obtain the Flipped model and by coupling the down-type
quarks to $\phi_2$ and the charged leptons to $\phi_1$
we obtain the Lepton Specific model.

We define
the signal strength as
\begin{equation}
\mu^{h_i}_f \, = \, \frac{\sigma \, {\rm BR} (h_i \to
  f)}{\sigma^{\scriptscriptstyle {\rm SM}} \, {\rm BR^{\scriptscriptstyle{\rm SM}}} (h_i \to f)}
\label{eg-rg}
\end{equation}
where $\sigma$ is the Higgs boson production cross section and ${\rm BR} (h_i \to f)$ is
the branching ratio of the $h_i$ decay into the final state $f$;  $\sigma^{\scriptscriptstyle {\rm {SM}}}$
and ${\rm BR^{\scriptscriptstyle {\rm SM}}}(h \to f)$ are the corresponding quantities calculated in the SM.
The cross sections were obtained from: HIGLU~\cite{Spira:1995mt} - gluon fusion at NNLO, 
together with the expressions for the CP-violating model in~\cite{Fontes:2014xva};
SusHi~\cite{Harlander:2012pb} - $b \bar{b} \ra h$ at NNLO; \cite{LHCCrossSections} - $Vh$ (associated production), 
$t \bar{t} h$ and $VV \ra h$ (vector boson fusion). 
The allowed parameter space of the C2HDM was recently reviewed in~\cite{Fontes:2015mea}
(see also~\cite{Fontes:2014xva, Ginzburg:2002wt, Khater:2003wq, ElKaffas:2007rq, ElKaffas:2006nt,
Grzadkowski:2009iz, Arhrib:2010ju, Barroso:2012wz}). The benchmark points that clearly signal CP-violation
will be presented in the next section and are chosen from this set. The allowed points in parameter
space are subject to the constraints we will briefly describe now. We note that we only focus here on scenarios
where the lightest scalar $h_1$ is the 125 GeV Higgs.

\begin{itemize}
\item
We take the lightest neutral scalar, $h_1$, to have a mass of 125 GeV in agreement
with the latest results from ATLAS~\cite{Aad:2014aba} and CMS~\cite{Khachatryan:2014jba}.
\item 
The accuracies in the measurements of the signal strengths in the processes $pp \to h_1 \to WW (ZZ)$, 
$pp \to h_1 \to \gamma \gamma$ and $pp \to h_1 \to \tau^+ \tau^-$ are about 
$20$\% at 1$\sigma$~\cite{Khachatryan:2014jba, ATLAS-CONF-2015-007}.
As shown in~\cite{Fontes:2014xva}, 
imposing these run 1 constraints guarantees that the C2HDM automatically obeys all other 
run 1 constraints on the 125 GeV Higgs decays in this model.
%at the end of run 1 these final states are enough to reproduce 
%quantitatively the effect of all possible final states in the Higgs decay in the C2HDM. 
We will 
thus force $\mu_{VV}$, $\mu_{\gamma \gamma}$ and $\mu_{\tau \tau}$ to be within $20$\%
of the expected SM value
\item
The LHC results also allow us to put bounds on the heavier scalars $h_2$ and $h_3$.
We impose the results on $\mu_{VV}$~\cite{Khachatryan:2015cwa} in the range $[145,1000]$
GeV and on $\mu_{\tau \tau}$~\cite{Khachatryan:2014wca} in the range $[100,1000]$
GeV. We also use the results on $h_i \to ZZ \to 4 l$ from~\cite{Chatrchyan:2013mxa}
in the range $[124,150]$ GeV and from \cite{Khachatryan:2015cwa} in the 
range $[150,990]$ GeV, and on $h \to \gamma \gamma$ from~\cite{Aad:2014ioa,Khachatryan:2014ira}.
Finally we also impose the constraints stemming from the results based on the searches 
$h_i \to Z h_1 \to Z b\bar b (\tau^+ \tau^-) $~\cite{Aad:2015wra} and  
$h_i \to Z h_1 \to ll b\bar b$~\cite{CMSHIG-14-011}.
\item
We consider the constraints on the charged Higgs Yukawa vertices that depend 
only on the charged Higgs mass and on $\tan \beta$. There is a new bound
on $b \ra s \gamma$, in Type II/F~\cite{Misiak:2015xwa} of $m_{H^\pm} \geq 480$ GeV
at 95\% C.L.. Putting together all the constraints from B-physics~\cite{Deschamps:2009rh, gfitter1} 
and also from the
$R_b\equiv\Gamma(Z\to b\bar{b})/\Gamma(Z\to{\rm hadrons})$~\cite{Ztobb} measurement, we can state
that roughly $\tan{\beta} \gtrsim 1$ for all models. LEP searches on $e^+ e^- \to H^+ H^-$~\cite{Abbiendi:2013hk}
and the LHC searches on $pp \to \bar t \, t (\to H^+ \bar b$~\cite{ATLASICHEP, CMSICHEP}) lead us to roughly 
consider $m_{H^\pm} \geq 100$ GeV in Type I/LS.
\item
We consider the following theoretical constraints: the potential has to be bounded from
below~\cite{Deshpande:1977rw}, perturbative unitarity is
required~\cite{Kanemura:1993hm, Akeroyd:2000wc,Ginzburg:2003fe} and all allowed points comply
with the oblique radiative parameters~\cite{Peskin:1991sw, Grimus:2008nb, Baak:2012kk}.
\item
The scenarios we will present in the next section are a clear signal of CP-violation in 
models with an extended scalar sector. Models with a CP-violating scalar sector
are constrained by bounds from electric dipole moments (EDMs) measurements. 
Although the search for the proposed final states should be performed from a model
independent perspective, we will nevertheless estimate the most important constraints
on the CP-violating phases in the context of the C2HDM~\cite{Buras:2010zm, Cline:2011mm, Jung:2013hka, 
Shu:2013uua, Inoue:2014nva, Brod:2013cka}.

The most stringent bound~\cite{Inoue:2014nva} comes from the ACME~\cite{Baron:2013eja} 
results on the ThO molecule EDM. In order to have points with EDMs of an order
of magnitude that conforms to the ACME result, we 
have computed the Barr-Zee diagrams with fermions in the loop. 
As we will see, the ACME bound can only be evaded by either going 
to the limit of the CP-conserving model or in scenarios where 
cancellations~\cite{Jung:2013hka, Shu:2013uua} among the neutral scalars occur. 
These cancellations are due to orthogonality of 
the $R$ matrix in the case of almost degenerate 
scalars~\cite{Fontes:2014xva}. 
We should finally point out that ref.~\cite{Jung:2013hka} 
argues that the extraction of the electron EDM from the data is filled with uncertainties
and an order of magnitude larger EDM than that claimed by ACME should be allowed 
for.
\end{itemize}

\section{CP-violating benchmark points}
\label{sec:bench}

 \begin{table}[!h]
 \small
 \centering
\begin{tabular}{|c|l|l|l|l|}\hline
 &$P1$&$P2$&$P3$&$P4$\\\hline
$\alpha_1$ &1.12569 &1.04842 & -1.33589 &1.41610\\
$\alpha_2$ &0.49091 & -0.00825 & -0.00129 &0.24037\\
$\alpha_3$ & -1.56775 &0.00674 &0.63749 & -0.81993\\
$\beta$ &0.92913 &1.00182 &1.27669 &1.29413\\
 $\tan\beta$ &1.33845 &1.56366 &3.30155 &3.52171\\
$m_1$ (GeV) &125.00 &125.00 &125.00 &125.00\\
$m_2$ (GeV) &127.32 &273.15 &282.53 &231.74\\
$m_3$ (GeV) &252.63 &421.64 &287.80 &360.59\\
 $m_{H^\pm}$ (GeV) &481.25 &452.50 &604.89 &527.67\\
 Re($m_{12}^2$) (GeV)${}^2$ &-0.5625E+02 & 0.1183E+05 & 0.1590E+05 & 0.2156E+05\\
 \hline
 \hline
 $b_{D_{1}}$ & -0.63099 &0.01291 &0.00426 & -0.83837\\
 $b_{L_{1}}$ & -0.63099 &0.01291 &0.00426 &0.06760\\
 \hline
{\bf\red $C_1$[1]} $\sigma_3\times $BR($h_3\to h_2 Z \to b\bar{b} l\bar{l}$) & 114.528\ [fb] &61.529\ [fb] &0.000\ [fb] &27.484\ [fb]\\
{\bf\red $C_1$[2]} $\sigma_2\times $BR($h_2\to h_1 Z\to b\bar{b} l\bar{l}$)&0.000\ [fb] &0.615\ [fb] &7.401\ [fb] &18.462\ [fb]\\
 {\bf\red$C_1$[3]} $\sigma_3\times $BR($h_3\to h_1 Z\to b\bar{b} l\bar{l}$) &26.656\ [fb] &1.100\ [fb] &24.519\ [fb] &1.787\ [fb]\\
 \hline
{\bf\red $C_2$[1]} $\sigma_2\times $BR($h_2\to h_1 Z\to b\bar{b} l\bar{l}$) &0.000\ [fb] &0.615\ [fb] &7.401\ [fb] &18.462\ [fb]\\
{\bf\red $C_2$[2]} $\sigma_1\times $BR($h_1\to Z Z\to l\bar{l} l\bar{l}$) &5.495\ [fb] &5.792\ [fb] &5.592\ [fb] &4.802\ [fb]\\
{\bf\red $C_2$[3]} $\sigma_2\times $BR($h_2\to Z Z\to l\bar{l} l\bar{l}$) &1.386\ [fb] &2.598\ [fb] &1.802\ [fb] &1.220\ [fb]\\
 \hline
{\bf\red $C_3$[1]} $\sigma_3\times $BR($h_3\to h_1 Z\to b\bar{b} l\bar{l}$) &26.656\ [fb] &1.100\ [fb] &24.519\ [fb] &1.787\ [fb]\\
{\bf\red $C_3$[2]} $\sigma_1\times $BR($h_1\to Z Z\to l\bar{l} l\bar{l}$) &5.495\ [fb] &5.792\ [fb] &5.592\ [fb] &4.802\ [fb]\\
{\bf\red $C_3$[3]} $\sigma_3\times $BR($h_3\to Z Z\to l\bar{l} l\bar{l}$) &1.011\ [fb] &0.003\ [fb] &1.733\ [fb] &1.058\ [fb]\\
 \hline
{\bf\red $C_4$[1]} $\sigma_3\times $BR($h_3\to h_2 Z\to b\bar{b} l\bar{l}$) & 114.528\ [fb] &61.529\ [fb] &0.000\ [fb] &27.484\ [fb]\\
{\bf\red $C_4$[2]} $\sigma_2\times $BR($h_2\to Z Z\to l\bar{l} l\bar{l}$) &1.386\ [fb] &2.598\ [fb] &1.802\ [fb] &1.220\ [fb]\\
{\bf\red $C_4$[3]} $\sigma_3\times $BR($h_3\to Z Z\to l\bar{l} l\bar{l}$) &1.011\ [fb] &0.003\ [fb] &1.733\ [fb] &1.058\ [fb]\\
 \hline
{\bf\red $C_5$[1]} $\sigma_3\times $BR($h_3\to Z Z\to l\bar{l} l\bar{l}$) &1.011\ [fb] &0.003\ [fb] &1.733\ [fb] &1.058\ [fb]\\
{\bf\red $C_5$[2]} $\sigma_2\times $BR($h_2\to Z Z\to l\bar{l} l\bar{l}$) &1.386\ [fb] &2.598\ [fb] &1.802\ [fb] &1.220\ [fb]\\
{\bf\red $C_5$[3]} $\sigma_1\times $BR($h_1\to Z Z\to l\bar{l} l\bar{l}$) &5.495\ [fb] &5.792\ [fb] &5.592\ [fb] &4.802\ [fb]\\
 \hline \end{tabular}
\caption{Benchmark points for Type II: $P1$, $P2$ and $P3$, and for the Flipped model: $P4$,
 for LHC at $\sqrt{s}=13$ TeV. All Z bosons decay leptonically which corresponds to a factor of $0.06732$
 for each Z decay.}
 \label{tab:bench1}
 \end{table}

In this section, we present some benchmark points that allow us to definitely
probe CP-violation during LHC's run 2. In table~\ref{tab:bench1}, we present four
benchmark points, where the first three are for Type II and $P4$ is for the Flipped
model (Type I and Lepton Specific are discussed in appendix A).
For each point we give the values of the parameters of the model, the values
of the pseudoscalar component of the Yukawa coupling of the lightest Higgs and the values of the cross sections
for the different processes.
The cross sections are calculated assuming that all scalars in the final state 
are detected in the decay to $b \bar b$ and all Z bosons are detected in the 
leptonic decays, providing therefore a very conservative estimate for the number of signal 
events available. Regarding the cross sections, we sum over all possible production process
with one scalar in the final state. Therefore, the numbers presented in the table correspond either to
\begin{equation}
\sigma (pp \to h_i + X) \, BR(h_i \to h_j Z) \, BR(h_j \to b \bar b) \, BR(Z \to ll) \, ,
\end{equation}
or
\begin{equation}
\sigma (pp \to h_i + X) \, BR(h_i \to Z Z) \, BR^2(Z \to ll)
\end{equation}
and $l=e, \, \mu$. 

The general criteria for the choice of our benchmark points is the following:
the points have passed all the constraints described in the previous section; the number 
of events for a luminosity of 100$fb^{-1}$ should be at least above 50, and the smallest
number in table~\ref{tab:bench1} for this luminosity is 61 events. Note that this number
already takes into account the decay of the scalar into $b \bar b$ and the decay of all
Z bosons into leptons (a reduction of $0.06732$ for each Z). 
Therefore, we expect a much larger number of events 
when all other combinations of final states are taken into
account by the experiments (as it is obviously the case for the $ZZ$
final states, where we can have combinations of leptons and jets final states).
\begin{table}[!h]
 \small
 \centering
  \begin{tabular}{|c|l|l|l|l|}\hline
 &$P1$&$P2$&$P3$&$P4$\\\hline
 $\mu_{WW}(h_1)=\mu_{ZZ}(h_1)$ &1.09016 &1.14962 &1.11696 &0.95402\\
  $\mu_{\tau\tau}(h_1)$ &1.16717 &0.98826 &0.96621 &1.02628\\
  $\mu_{\gamma\gamma}(h_1)$ &0.92139 &1.02589 &0.87922 &0.85345\\
 $\mu_{bb(VH)}(h_1)$ &0.71662 &0.93593 &0.65922 &0.94294\\
 \hline
$\mu_{WW}(h_2)/\mu_{WW}^{\rm exp}$&0.225/NA\, &0.151/0.185&0.117/0.170&0.058/0.121\\
$\mu_{ZZ}(h_2)/\mu_{ZZ}^{\rm exp}$&0.225/1.264&0.151/0.190&0.117/0.176&0.058/0.130\\
  $\mu_{\tau\tau}(h_2)/\mu_{\tau\tau}^{\rm exp}$& 1.59/ 3.98&   180.00/   472.37& 7.98/   490.42& 0.90/   363.88\\
  $\sigma BR_{\gamma\gamma}(h_2)/\sigma BR_{\gamma\gamma}^{\rm exp}$ [fb]&   15.265/   29.705&0.318/2.678&0.011/2.727&0.018/5.998\\
  $\mu_{\gamma\gamma}(h_2)/\mu_{\gamma\gamma}^{\rm exp}$ ($m<150$GeV)&0.258/0.259&0.000/0.000&0.000/0.000&0.000/0.000\\
 $\sigma BR_{Zh\to Zbb}(h_2)/\sigma BR_{Zh\to Zbb}^{\rm exp}$ [pb]&0.000/   0.000&0.003/0.308&0.042/0.250&0.108/0.403\\
   $\sigma BR_{Zh\to Z\tau\tau}(h_2)/\sigma BR_{Zh\to Z\tau\tau}^{\rm exp}$ [pb]&0.000/   0.000&0.000/0.105&0.005/0.089&0.012/0.085\\
$\sigma BR_{Zh\to ll bb}(h_2)/\sigma BR_{Zh\to ll bb}^{\rm exp}$ [fb]&0.000/0.000&0.222/   15.242&2.855/   12.167&7.259/   14.082\\
 \hline
$\mu_{WW}(h_3)/\mu_{WW}^{\rm exp}$&0.053/0.074&0.000/0.083&0.111/0.125&0.072/0.099\\
$\mu_{ZZ}(h_3)/\mu_{ZZ}^{\rm exp}$&0.053/0.068&0.000/0.086&0.111/0.147&0.072/0.095\\
  $\mu_{\tau\tau}(h_3)/\mu_{\tau\tau}^{\rm exp}$& 3.12/   427.59& 8.70/  1241.83&13.52/   500.43& 0.04/   663.64\\
  $\sigma BR_{\gamma\gamma}(h_3)/\sigma BR_{\gamma\gamma}^{\rm exp}$ [fb]&0.022/6.511&0.028/2.002&0.010/2.672&0.004/2.823\\
 $\sigma BR_{Zh\to Zbb}(h_3)/\sigma BR_{Zh\to Zbb}^{\rm exp}$ [pb]&0.147/0.310&0.005/0.081&0.135/0.228&0.009/0.156\\
   $\sigma BR_{Zh\to Z\tau\tau}(h_3)/\sigma BR_{Zh\to Z\tau\tau}^{\rm exp}$ [pb]&0.017/0.102&0.001/0.035&0.016/0.081&0.001/0.038\\
$\sigma BR_{Zh\to ll bb}(h_3)/\sigma BR_{Zh\to ll bb}^{\rm exp}$ [fb]&9.926/   23.839&0.337/2.731&9.076/   15.230&0.605/7.358\\
 \hline \end{tabular}
 \label{tab:bench2}
 \caption{Constraints from the LHC at $\sqrt{s}=8$ TeV for the benchmark points $P1$,
 $P2$ and $P3$ (Type II) and $P4$ (Flipped). NA stands for not available.}
 \end{table}
In table~\ref{tab:bench2} we show the rates obtained for the benchmark points which are then compared
to the available experimental data from the LHC at $\sqrt{s}=8$ TeV. 

 \begin{table}[!h]
 \small
 \centering
  \begin{tabular}{|c|l|l|l|l|}\hline
 &$P1$&$P2$&$P3$&$P4$\\\hline
 \hline
   $\sigma(h_1)$ {\bf 13TeV} &61.600  \ [pb] &53.217  \ [pb] &54.825  \ [pb] &51.275  \ [pb]\\
$\sigma(h_1)$BR($h_1\to W^*W^*$) &11.819  \ [pb] &12.459  \ [pb] &12.028  \ [pb] &10.328  \ [pb]\\
$\sigma(h_1)$BR($h_1\to Z^*Z^*$) & 1.212  \ [pb] & 1.278  \ [pb] & 1.234  \ [pb] & 1.060  \ [pb]\\
$\sigma(h_1)$BR($h_1\to bb$) &34.383  \ [pb] &29.087  \ [pb] &28.256  \ [pb] &30.313  \ [pb]\\
  $\sigma(h_1)$BR($h_1\to \tau\tau$) & 3.969  \ [pb] & 3.360  \ [pb] & 3.264  \ [pb] & 3.485  \ [pb]\\
  $\sigma(h_1)$BR($h_1\to \gamma\gamma$) &   129.973  \ [fb] &   144.664  \ [fb] &   123.188  \ [fb] &   120.222  \ [fb]\\
 \hline
 \hline
 $\sigma_2\equiv\sigma(h_2)$ {\bf 13TeV} &56.583  \ [pb] & 4.262  \ [pb] & 1.602  \ [pb] & 3.354  \ [pb]\\
 $\sigma_2\times$BR($h_2\to WW$) & 2.814  \ [pb] & 1.323  \ [pb] & 0.910  \ [pb] & 0.656  \ [pb]\\
$\sigma_2\times$BR($h_2\to ZZ$)  & 0.306  \ [pb] & 0.573  \ [pb] & 0.398  \ [pb] & 0.269  \ [pb]\\
 $\sigma_2\times$BR($h_2\to bb$) &42.534  \ [pb] & 1.894  \ [pb] & 0.067  \ [pb] & 1.944  \ [pb]\\
   $\sigma_2\times$BR($h_2\to \tau\tau$) & 4.911  \ [pb] & 0.224  \ [pb] & 0.008  \ [pb] & 0.002  \ [pb]\\
   $\sigma_2\times$BR($h_2\to \gamma\gamma$) &35.041  \ [fb] & 0.879  \ [fb] & 0.027  \ [fb] & 0.046  \ [fb]\\
 \hline
  $\sigma_2\times$BR($h_2\to h_1 Z$) & 0.000  \ [pb] & 0.017  \ [pb] & 0.213  \ [pb] & 0.464  \ [pb]\\
  $\sigma_2\times$BR($h_2\to h_1 Z\to bb Z$) & 0.000  \ [pb] & 0.009  \ [pb] & 0.110  \ [pb] & 0.274  \ [pb]\\
$\sigma_2\times$BR($h_2\to h_1 Z\to \tau\tau Z$) & 0.000  \ [fb] & 1.055  \ [fb] &12.697  \ [fb] &31.530  \ [fb]\\
 \hline
$\sigma_2\times$BR($h_2\to h_1 h_1$) & 0.000  \ [fb] & 0.007  \ [fb] & 5.016  \ [fb] & 0.000  \ [fb]\\
  $\sigma_2\times$BR($h_2\to h_1 h_1\to bb\ bb$) & 0.000  \ [fb] & 0.002  \ [fb] & 1.332  \ [fb] & 0.000  \ [fb]\\
$\sigma_2\times$BR($h_2\to h_1 h_1\to bb\ \tau\tau$) & 0.000  \ [fb] & 0.000  \ [fb] & 0.308  \ [fb] & 0.000  \ [fb]\\
  $\sigma_2\times$BR($h_2\to h_1 h_1\to \tau\tau\ \tau\tau$) & 0.000  \ [fb] & 0.000  \ [fb] & 0.018  \ [fb] & 0.000  \ [fb]\\
 \hline
 \hline
 $\sigma_3\equiv\sigma(h_3)$ {\bf 13TeV} & 4.043  \ [pb] & 8.480  \ [pb] & 2.086  \ [pb] & 1.819  \ [pb]\\
 $\sigma_3\times$BR($h_3\to WW$) & 0.526  \ [pb] & 0.001  \ [pb] & 0.871  \ [pb] & 0.509  \ [pb]\\
 $\sigma_3\times$BR($h_3\to ZZ$) & 0.223  \ [pb] & 0.001  \ [pb] & 0.382  \ [pb] & 0.233  \ [pb]\\
 $\sigma_3\times$BR($h_3\to bb$) & 0.047  \ [pb] & 0.016  \ [pb] & 0.109  \ [pb] & 0.058  \ [pb]\\
   $\sigma_3\times$BR($h_3\to \tau\tau$) & 5.558  \ [fb] & 1.913  \ [fb] &12.856  \ [fb] & 0.020  \ [fb]\\
  $\sigma_3\times$BR($h_3\to \gamma\gamma$)  & 0.059  \ [fb] & 0.093  \ [fb] & 0.028  \ [fb] & 0.013  \ [fb]\\
 \hline
  $\sigma_3\times$BR($h_3\to h_1 Z$) & 0.709  \ [pb] & 0.030  \ [pb] & 0.707  \ [pb] & 0.045  \ [pb]\\
  $\sigma_3\times$BR($h_3\to h_1 Z\to bb Z$) & 0.396  \ [pb] & 0.016  \ [pb] & 0.364  \ [pb] & 0.027  \ [pb]\\
$\sigma_3\times$BR($h_3\to h_1 Z\to \tau\tau Z$) &45.708  \ [fb] & 1.887  \ [fb] &42.067  \ [fb] & 3.051  \ [fb]\\
 \hline
  $\sigma_3\times$BR($h_3\to h_2 Z$) & 2.263  \ [pb] & 2.057  \ [pb] & 0.000  \ [pb] & 0.705  \ [pb]\\
  $\sigma_3\times$BR($h_3\to h_2 Z\to bb Z$) & 1.701  \ [pb] & 0.914  \ [pb] & 0.000  \ [pb] & 0.408  \ [pb]\\
$\sigma_3\times$BR($h_3\to h_2 Z\to \tau\tau Z$) &   196.416  \ [fb] &   107.996  \ [fb] & 0.000  \ [fb] & 0.500  \ [fb]\\
 \hline
$\sigma_3\times$BR($h_3\to h_1 h_1$) & 0.090  \ [fb] & 0.230  \ [fb] & 2.071  \ [fb] &19.918  \ [fb]\\
  $\sigma_3\times$BR($h_3\to h_1 h_1\to bb\ bb$) & 0.028  \ [fb] & 0.069  \ [fb] & 0.550  \ [fb] & 6.961  \ [fb]\\
$\sigma_3\times$BR($h_3\to h_1 h_1\to bb\ \tau\tau$) & 0.007  \ [fb] & 0.016  \ [fb] & 0.127  \ [fb] & 1.601  \ [fb]\\
  $\sigma_3\times$BR($h_3\to h_1 h_1\to \tau\tau\ \tau\tau$) & 0.000  \ [fb] & 0.001  \ [fb] & 0.007  \ [fb] & 0.092  \ [fb]\\
 \hline
$\sigma_3\times$BR($h_3\to h_2 h_1$) &   263.916  \ [fb] & 0.038  \ [fb] & 0.000  \ [fb] &11.157  \ [fb]\\
  $\sigma_3\times$BR($h_3\to h_2 h_1\to bb\ bb$) &   110.732  \ [fb] & 0.009  \ [fb] & 0.000  \ [fb] & 3.822  \ [fb]\\
$\sigma_3\times$BR($h_3\to h_2 h_1\to bb\ \tau\tau$) &25.567  \ [fb] & 0.002  \ [fb] & 0.000  \ [fb] & 0.444  \ [fb]\\
  $\sigma_3\times$BR($h_3\to h_2 h_1\to \tau\tau\ \tau\tau$) & 1.476  \ [fb] & 0.000  \ [fb] & 0.000  \ [fb] & 0.001  \ [fb]\\
 \hline \end{tabular}
 \caption{Predictions for $\sigma \times  {\rm BR}$
  for the LHC at $\sqrt{s}=13$ TeV for the
 benchmark points $P1$, $P2$ and $P3$ (Type II) and $P4$
  (Flipped).}
 \label{tab:bench3}
 \end{table}

The criteria for the choice of each particular point is severely constrained by the ACME results.
In fact, all the points have similar features in that they either have two neutral scalar masses
almost degenerate or values of the angles very close to zero (therefore approaching the limit of the
CP-conserving 2HDM). Points $P1$ and $P3$ have degenerate masses while point $P2$ has very small $\alpha_2$
and $\alpha_3$ values. That is why for point $P2$, the decay $h_2 \to h_1 Z$ is suppressed. In the
limit $\alpha_2 = \alpha_3 = 0$, $h_3$ is the pseudo-scalar and $h_1$ and $h_2$ are scalars and  
$h_2 \to h_1 Z$ is forbidden. For the same reason, $h_3 \to ZZ$ is forbidden. 
Note however that although $\alpha_2$ and $\alpha_3$ are very small 
we still have a large number of signal events for 100$fb^{-1}$ in $h_2 \to h_1 Z$.
As $\alpha_{2,3}$ move away from zero (the CP-conserving limit) certain CP-violating
observables grow extremely fast. Thus, we can be very close to this limit
and still have large CP-violating signals.

The points were also chosen so that they would probe more than one class simultaneously. $P1$
probes classes $C_3$, $C_4$ and $C_5$; $P2$ probes $C_1$ and $C_2$; $P3$ probes $C_2$, $C_3$ and $C_5$
while the point for the Flipped model probes all classes. Furthermore, points $P1$ and $P4$
were also chosen to show that large pseudoscalar components are not only still
allowed, as previously discussed in~\cite{Fontes:2015mea}, but they can also easily be probed at the 
next LHC run.

Finally, in table~\ref{tab:bench3} we present the production cross sections for $h_1$, $h_2$
and $h_3$. In the same table we show the $\sigma (h_i) \times Br(h_i \to X)$ where $X$ stands
for the main final sates being searched by ATLAS and CMS at the next LHC run. These numbers
allow the experimental groups to understand if a given scalar is found in direct production
whether it comes from a CP-violating process or not. In the same table we also present 
the values of the scalar production cross sections that lead to decays of the type $h_i \to h_j h_j$
and $h_i \to h_j h_k$ and that are clearly too small to be detected at the LHC for the sets of 
benchmark chosen, except for a few cases for points $P1$ and $P4$.

\subsection{CP-violating scenarios involving scalar to two scalars decays}
\label{subsec:cp2}

There are other classes of decays that constitute a sign of CP-violation in the 2HDM.
Some of them involve the decay $h_3 \to h_2 h_1$ which is not present in the CP-conserving
version of the 2HDM. In fact the decay $h_3 \to h_2 h_1$ is only possible if either
all $h_i$ are CP-even, two of the $h_i$ are CP-odd and one is CP-even or if CP
is not conserved. Since the decay $h_1 \to ZZ$ was already observed we know $h_1$
has a CP-even component. Therefore, we can discuss the combinations of decays that
together with $h_3 \to h_2 h_1$ will be a clear sign of CP-violation in the 2HDM
or that will point to other extensions of the SM that can be either CP-conserving
or CP-violating.  

In table~\ref{tab:classes2} we present new classes of decays that again constitute
model independent signs of CP-violation. Class $C_6$ is composed 
by the three decays $h_1 \rightarrow Z Z$, $h_3 \rightarrow h_2 h_1$
and $h_3\to h_2 Z$.
There are other sets of simultaneous measurements
involving $h_3 \rightarrow h_2 h_1$ that are consistent with CP conservation in models with
more than two Higgs doublets, and which allow the determination of the possible CP
assignments. These are:
\begin{itemize}
\item $[h_1 \rightarrow Z Z;\ h_3 \rightarrow h_2 h_1]$ $h_3 \rightarrow h_1 
Z$,
leading to the CP assignments (+, $-$, $-$);
\item $[h_1 \rightarrow Z Z;\ h_3 \rightarrow h_2 h_1]$ $h_2 \rightarrow h_1 
Z$,
leading to the CP assignments (+, $-$, $-$);
\item $[h_1 \rightarrow Z Z;\ h_3 \rightarrow h_2 h_1]$ $h_2 \rightarrow Z 
Z$,
leading to the CP assignments (+, +, +).
\end{itemize}
There are still other combinations involving scalar to scalar decays
that are model independent signs of CP-violation. Such is the case 
of class $C_7$ in table~\ref{tab:classes2} composed by the decays
$h_{2,3} \to h_1 h_1$, $h_{2,3} \to h_1 Z$ and $h_1 \to ZZ$.
Finally, other combinations like $h_3 \to h_1 h_1 (h_2 h_2)$, $h_2 \to h_1 h_1$ and
$h_1 \to ZZ$ are not possible in a CP-conserving 2HDM but are possible in the C2HDM
and can also serve to determine the CP-quantum numbers of other extensions of the
scalar sector.
A detailed study of these classes will be performed
in a forthcoming publication~\cite{us}.

\begin{table}[!h]
 \small
 \centering
\begin{tabular}{|c|l|l|l|l|l|}\hline
Classes & $C_6$ & $C_7$ \\\hline
        & $h_3\to h_2 h_1$  & $h_{2,3} \to h_1 h_1$ \\
Decays  & $h_3\to h_2 Z$  & $h_{2,3} \to h_1 Z$ \\
        & $h_1\to Z Z$ & $h_1\to Z Z$ \\
 \hline \end{tabular}
 \caption{\label{tab:classes2}Classes of combined measurements guaranteed
 to probe CP-violation.}
\end{table}

\section{Conclusions}
\label{sec:concl}

We have proposed five classes of processes that constitute conclusive evidence 
of CP-violation in scalar decays. While the $C_5$ class is particular to the C2HDM,
all other classes are valid in any scalar extension of the SM. One of the most attractive 
features of our proposal is to rely on searches that are already planned for the
LHC run 2, namely $h_i \to ZZ$ and $h_i \to h_j Z$. Furthermore, it does not depend
on complex distributions nor asymmetries of any kind, but only on total rates of specific processes.
It is a direct and straightforward way to search for CP-violation at the LHC in scalar
decays. As far as we know this is the only method of probing CP-violation based on rates
only.

We have shown that even taking into account all constraints and in particular the one
from ACME that heavily restricts the amount of CP-violation in the model, it is still
easy to find points to probe each of the proposed classes. In many cases a point
can be used to probe several classes simultaneously. We have chosen a set of benchmark
points according to different criteria, always keeping in mind that the decays should 
be within the reach of the LHC's run 2. We should point out however that even if these
points are excluded the parameter space is large enough to provide many more points
and the model is far from excluded (nor is CP-violation in scalar decays excluded).
The future bounds on EDMs\cite{Brod:2013cka, Dekens:2014jka} can have a strong impact
on the allowed parameter space, and one has to consider the interplay between 
the EDM bounds and the data from run 2 to propose scenarios for future experiments.
However, the EDM constraints get looser if one goes beyond the setting discussed here,
allowing for $\lambda_6 \neq 0$ and/or $\lambda_7 \neq 0$~\cite{Chen:2015gaa}. In that
case classes $C_1$ to $C_5$ still probe CP-violation, and thus the methods proposed
here should be pursued experimentally regardless of the fate of the C2HDM. 
In particular, classes $C_1$ to $C_4$ probe CP-violation in all models.

We also propose two new classes of decays, $C_6$ and $C_7$ that involve the already
observed decay $h_1 \to ZZ$, one decay of the type $h_i \to h_j h_{j(k)}$
with $j \neq k$ and one decay of type $h_j \to h_k Z$.

As important guidelines for experiments, we propose six benchmark points covering
all C2HDM types: type II ($P1$-$P3$), and Flipped ($P4$) in tables~\ref{tab:bench1},
\ref{tab:bench2} and \ref{tab:bench3}; type I ($P5$) and Lepton Specific ($P6$) in 
tables~\ref{tab:bench4}, \ref{tab:bench5} and \ref{tab:bench6} of appendix A.
We provide all event rates for all scalar processes and for each benchmark points.
This allows not only to search for the CP-violating classes of decays but also
to confirm or disprove the points via direct search of each scalar. If a particular
point is found the other decays could clarify if we are in presence of the C2HDM
or of some other CP-violating extension of the SM.

\appendix
\section{Benchmark points for Type I and for the Lepton Specific models}
\label{sec:app}

\begin{table}[!h]
 \small
 \centering
  \begin{tabular}{|c|l|l|}\hline
 &$P5$&$P6$\\\hline
$\alpha_1$ & 1.30680 & 1.08742\\
$\alpha_2$ & 0.10867 & 0.00960\\
$\alpha_3$ &-0.20624 &-0.41962\\
$\beta$ & 1.15333 & 1.03051\\
  $\tan\beta$ & 2.25459 & 1.66717\\
 $m_1$ (GeV) & 125.00 & 125.00\\
 $m_2$ (GeV) & 235.45 & 262.98\\
 $m_3$ (GeV) & 359.20 & 264.60\\
 $m_{H^\pm}$ (GeV) & 522.87 & 471.76\\
  Re($m_{12}^2$) (GeV)${}^2$ &0.9504E+02 &  -0.3006E+05\\
 \hline
 \hline
  $b_{D_{1}}$ & 0.04810 & 0.00576\\
  $b_{L_{1}}$ & 0.04810 &-0.01600\\
 \hline
 {\bf\red $C_1$[1]} $\sigma_3\times $BR($h_3\to h_2 Z \to b\bar{b} l\bar{l}$) &  1.251  \ [fb] &  0.000  \ [fb]\\
 {\bf\red $C_1$[2]} $\sigma_2\times $BR($h_2\to h_1 Z\to b\bar{b} l\bar{l}$)  &  5.644  \ [fb] &  3.030  \ [fb]\\
  {\bf\red $C_1$[3]} $\sigma_3\times $BR($h_3\to h_1 Z\to b\bar{b} l\bar{l}$) & 15.477  \ [fb] & 27.984  \ [fb]\\
 \hline
  {\bf\red $C_2$[1]} $\sigma_2\times $BR($h_2\to h_1 Z\to b\bar{b} l\bar{l}$) &  5.644  \ [fb] &  3.030  \ [fb]\\
 {\bf\red $C_2$[2]} $\sigma_1\times $BR($h_1\to Z Z\to l\bar{l} l\bar{l}$) &  4.954  \ [fb] &  5.146  \ [fb]\\
 {\bf\red $C_2$[3]} $\sigma_2\times $BR($h_2\to Z Z\to l\bar{l} l\bar{l}$) &  1.934  \ [fb] &  1.053  \ [fb]\\
 \hline
  {\bf\red $C_3$[1]} $\sigma_3\times $BR($h_3\to h_1 Z\to b\bar{b} l\bar{l}$) & 15.477  \ [fb] & 27.984  \ [fb]\\
 {\bf\red $C_3$[2]} $\sigma_1\times $BR($h_1\to Z Z\to l\bar{l} l\bar{l}$) &  4.954  \ [fb] &  5.146  \ [fb]\\
 {\bf\red $C_3$[3]} $\sigma_3\times $BR($h_3\to Z Z\to l\bar{l} l\bar{l}$) &  1.326  \ [fb] &  1.840  \ [fb]\\
 \hline
  {\bf\red $C_4$[1]} $\sigma_3\times $BR($h_3\to h_2 Z\to b\bar{b} l\bar{l}$) &  1.251  \ [fb] &  0.000  \ [fb]\\
 {\bf\red $C_4$[2]} $\sigma_2\times $BR($h_2\to Z Z\to l\bar{l} l\bar{l}$) &  1.934  \ [fb] &  1.053  \ [fb]\\
 {\bf\red $C_4$[3]} $\sigma_3\times $BR($h_3\to Z Z\to l\bar{l} l\bar{l}$) &  1.326  \ [fb] &  1.840  \ [fb]\\
 \hline
 {\bf\red $C_5$[1]} $\sigma_3\times $BR($h_3\to Z Z\to l\bar{l} l\bar{l}$) &  1.326  \ [fb] &  1.840  \ [fb]\\
 {\bf\red $C_5$[2]} $\sigma_2\times $BR($h_2\to Z Z\to l\bar{l} l\bar{l}$) &  1.934  \ [fb] &  1.053  \ [fb]\\
 {\bf\red $C_5$[3]} $\sigma_1\times $BR($h_1\to Z Z\to l\bar{l} l\bar{l}$) &  4.954  \ [fb] &  5.146  \ [fb]\\
 \hline \end{tabular}  
\caption{Benchmark points for Type I: $P5$ and for the Lepton Specific model: $P6$,
 for LHC at $\sqrt{s}=13$ TeV. All Z decay leptonically corresponding to a factor of $0.06732$.}
 \label{tab:bench4} 
 \end{table}  

In this appendix we present two further benchmark points, one for Type I and the other
for the Lepton Specific (LS) model. In table~\ref{tab:bench4} we present the values
of the parameters and the cross sections for benchmark point $P5$ in Type I
and  $P6$ for the LS model. In Type I it was possible to find a point that
not only complies with all the constraints but that probes all CP-violating
classes at the same time. For the LS model the classes probed are $C_2$, $C_3$
and $C_5$.

 \begin{table}[!h]
 \small
 \centering
  \begin{tabular}{|c|l|l|}\hline
 &$P5$&$P6$\\\hline
 $\mu_{WW}(h_1)=\mu_{ZZ}(h_1)$ &0.98240 &1.02070\\
  $\mu_{\tau\tau}(h_1)$ &1.12419 &0.83628\\
  $\mu_{\gamma\gamma}(h_1)$ &0.84875 &0.86872\\
 $\mu_{bb(VH)}(h_1)$ &0.99480 &1.02881\\
 \hline
$\mu_{WW}(h_2)/\mu_{WW}^{\rm exp}$&0.091/0.115&0.058/0.108\\
$\mu_{ZZ}(h_2)/\mu_{ZZ}^{\rm exp}$&0.091/0.111&0.058/0.112\\
  $\mu_{\tau\tau}(h_2)/\mu_{\tau\tau}^{\rm exp}$& 0.56/   377.80&72.97/   451.42\\
  $\sigma BR_{\gamma\gamma}(h_2)/\sigma BR_{\gamma\gamma}^{\rm exp}$ [fb]&0.046/3.975&0.125/6.838\\
  $\mu_{\gamma\gamma}(h_2)/\mu_{\gamma\gamma}^{\rm exp}$ ($m<150$GeV)&0.000/0.000&0.000/0.000\\
 $\sigma BR_{Zh\to Zbb}(h_2)/\sigma BR_{Zh\to Zbb}^{\rm exp}$ [pb]&0.032/0.337&0.016/0.349\\
   $\sigma BR_{Zh\to Z\tau\tau}(h_2)/\sigma BR_{Zh\to Z\tau\tau}^{\rm exp}$ [pb]&0.004/0.080&0.001/0.114\\
$\sigma BR_{Zh\to ll bb}(h_2)/\sigma BR_{Zh\to ll bb}^{\rm exp}$ [fb]&2.127/   13.013&1.100/   27.341\\
 \hline
$\mu_{WW}(h_3)/\mu_{WW}^{\rm exp}$&0.087/0.097&0.102/0.113\\
$\mu_{ZZ}(h_3)/\mu_{ZZ}^{\rm exp}$&0.087/0.094&0.102/0.123\\
  $\mu_{\tau\tau}(h_3)/\mu_{\tau\tau}^{\rm exp}$& 0.89/   656.23&   281.79/   454.89\\
  $\sigma BR_{\gamma\gamma}(h_3)/\sigma BR_{\gamma\gamma}^{\rm exp}$ [fb]&0.046/2.758&0.875/6.334\\
 $\sigma BR_{Zh\to Zbb}(h_3)/\sigma BR_{Zh\to Zbb}^{\rm exp}$ [pb]&0.075/0.155&0.151/0.348\\
   $\sigma BR_{Zh\to Z\tau\tau}(h_3)/\sigma BR_{Zh\to Z\tau\tau}^{\rm exp}$ [pb]&0.009/0.038&0.013/0.114\\
$\sigma BR_{Zh\to ll bb}(h_3)/\sigma BR_{Zh\to ll bb}^{\rm exp}$ [fb]&5.077/7.483&   10.163/   24.919\\
 \hline \end{tabular}
 \caption{Constraints from the LHC at $\sqrt{s}=8$ TeV for the benchmark points 
 $P5$ (Type I) and $P6$ (Lepton Specific).}
\label{tab:bench5}
 \end{table}

As we did for the remaining benchmark points, we present in table~\ref{tab:bench5}
the effect of the  LHC constraints on the processes involving scalars. In table~\ref{tab:bench6}
we present the production cross sections for $h_1$, $h_2$
and $h_3$ and also the $\sigma (h_i) \times Br(h_i \to X)$ where again $X$ stands
for the most relevant final states searched by ATLAS and CMS at the next LHC run. We also show 
the values of the scalar production cross sections that lead to decays of the type $h_i \to h_j h_j$
and $h_i \to h_j h_k$. Interestingly, for the benchmark points of Type I and Lepton Specific, there
are many scalar to scalar decays that could be probed at the next LHC run.

\begin{table}[!h]
 \small
 \centering
  \begin{tabular}{|c|l|l|}\hline
 &$P5$&$P6$\\\hline
 \hline
  $\sigma(h_1)$ {\bf 13TeV} & 55.144  \ [pb] & 53.455  \ [pb]\\
 $\sigma(h_1)$BR($h_1\to W^*W^*$) & 10.657  \ [pb] & 11.069  \ [pb]\\
 $\sigma(h_1)$BR($h_1\to Z^*Z^*$) &  1.093  \ [pb] &  1.136  \ [pb]\\
  $\sigma(h_1)$BR($h_1\to bb$) & 33.118  \ [pb] & 32.152  \ [pb]\\
  $\sigma(h_1)$BR($h_1\to \tau\tau$) &  3.825  \ [pb] &  2.845  \ [pb]\\
 $\sigma(h_1)$BR($h_1\to \gamma\gamma$) &119.794  \ [fb] &122.579  \ [fb]\\
 \hline
 \hline
$\sigma_2\equiv\sigma(h_2)$ {\bf 13TeV} &  1.620  \ [pb] &  4.920  \ [pb]\\
  $\sigma_2\times$BR($h_2\to WW$) &  1.032  \ [pb] &  0.542  \ [pb]\\
 $\sigma_2\times$BR($h_2\to ZZ$)  &  0.427  \ [pb] &  0.232  \ [pb]\\
  $\sigma_2\times$BR($h_2\to bb$) &  0.012  \ [pb] &  0.097  \ [pb]\\
  $\sigma_2\times$BR($h_2\to \tau\tau$) &  0.001  \ [pb] &  0.109  \ [pb]\\
 $\sigma_2\times$BR($h_2\to \gamma\gamma$) &  0.123  \ [fb] &  0.344  \ [fb]\\
 \hline
  $\sigma_2\times$BR($h_2\to h_1 Z$) &  0.140  \ [pb] &  0.075  \ [pb]\\
$\sigma_2\times$BR($h_2\to h_1 Z\to bb Z$) &  0.084  \ [pb] &  0.045  \ [pb]\\
$\sigma_2\times$BR($h_2\to h_1 Z\to \tau\tau Z$) &  9.683  \ [fb] &  3.982  \ [fb]\\
 \hline
$\sigma_2\times$BR($h_2\to h_1 h_1$) &  0.000  \ [fb] &  3772.577  \ [fb]\\
  $\sigma_2\times$BR($h_2\to h_1 h_1\to bb\ bb$) &  0.000  \ [fb] &  1364.787  \ [fb]\\
  $\sigma_2\times$BR($h_2\to h_1 h_1\to bb\ \tau\tau$) &  0.000  \ [fb] &241.505  \ [fb]\\
  $\sigma_2\times$BR($h_2\to h_1 h_1\to \tau\tau\ \tau\tau$) &  0.000  \ [fb] & 10.684  \ [fb]\\
 \hline
 \hline
$\sigma_3\equiv\sigma(h_3)$ {\bf 13TeV} &  9.442  \ [pb] & 10.525  \ [pb]\\
  $\sigma_3\times$BR($h_3\to WW$) &  0.638  \ [pb] &  0.945  \ [pb]\\
  $\sigma_3\times$BR($h_3\to ZZ$) &  0.293  \ [pb] &  0.406  \ [pb]\\
  $\sigma_3\times$BR($h_3\to bb$) &  0.004  \ [pb] &  0.422  \ [pb]\\
  $\sigma_3\times$BR($h_3\to \tau\tau$) &  0.432  \ [fb] &407.337  \ [fb]\\
$\sigma_3\times$BR($h_3\to \gamma\gamma$)  &  0.140  \ [fb] &  2.410  \ [fb]\\
 \hline
  $\sigma_3\times$BR($h_3\to h_1 Z$) &  0.383  \ [pb] &  0.691  \ [pb]\\
$\sigma_3\times$BR($h_3\to h_1 Z\to bb Z$) &  0.230  \ [pb] &  0.416  \ [pb]\\
$\sigma_3\times$BR($h_3\to h_1 Z\to \tau\tau Z$) & 26.554  \ [fb] & 36.779  \ [fb]\\
 \hline
  $\sigma_3\times$BR($h_3\to h_2 Z$) &  2.495  \ [pb] &  0.000  \ [pb]\\
$\sigma_3\times$BR($h_3\to h_2 Z\to bb Z$) &  0.019  \ [pb] &  0.000  \ [pb]\\
$\sigma_3\times$BR($h_3\to h_2 Z\to \tau\tau Z$) &  2.188  \ [fb] &  0.000  \ [fb]\\
 \hline
$\sigma_3\times$BR($h_3\to h_1 h_1$) &433.402  \ [fb] &  6893.255  \ [fb]\\
  $\sigma_3\times$BR($h_3\to h_1 h_1\to bb\ bb$) &156.329  \ [fb] &  2493.740  \ [fb]\\
  $\sigma_3\times$BR($h_3\to h_1 h_1\to bb\ \tau\tau$) & 36.111  \ [fb] &441.277  \ [fb]\\
  $\sigma_3\times$BR($h_3\to h_1 h_1\to \tau\tau\ \tau\tau$) &  2.085  \ [fb] & 19.521  \ [fb]\\
 \hline
$\sigma_3\times$BR($h_3\to h_2 h_1$) &  0.000  \ [fb] &  0.000  \ [fb]\\
  $\sigma_3\times$BR($h_3\to h_2 h_1\to bb\ bb$) &  0.000  \ [fb] &  0.000  \ [fb]\\
  $\sigma_3\times$BR($h_3\to h_2 h_1\to bb\ \tau\tau$) &  0.000  \ [fb] &  0.000  \ [fb]\\
  $\sigma_3\times$BR($h_3\to h_2 h_1\to \tau\tau\ \tau\tau$) &  0.000  \ [fb] &  0.000  \ [fb]\\
 \hline \end{tabular}
  \caption{Predictions for $\sigma \times  {\rm BR}$
  at $\sqrt{s}=13$ TeV for the
 benchmark points $P5$ (Type I) and $P6$
  (Lepton Specific).}
 \label{tab:bench6}
 \end{table}

%%%%%%%%%%%%%%%%%%%%%%%%%%%%%%%%%%%%%%%%%%%%%%%%%%%%%%%
\vspace*{0.5cm}
\section*{Acknowledgments}
J.C.R. and J.P.S. are supported in part by the Portuguese
\textit{Funda\c{c}\~{a}o para a Ci\^{e}ncia e Tecnologia}
under contract UID/FIS/00777/2013.
We acknowledge discussions with  Nikolaos Rompotis and Andr\'e David
regarding the ATLAS and CMS experimental analysis
respectively. We also acknowledge discussions with Augusto Barroso, Francisco Botella and Lu\'{\i}s Lavoura. 
\vspace*{0.5cm}
%%%%%%%%%%%%%%%%%%%%%%%%%%%%%%%%%%%%%%%%%%%%%%%%%%%%%%%

\end{document}